*Review*

# A Survey on Gas Sensing Technology

**Xiao Liu [1], Sitian Cheng [1], Hong Liu [1], Sha Hu [1], Daqiang Zhang [2] and Huansheng Ning [1],***

[1] School of Electronic and Information Engineering, Beihang University, Beijing 100191, China;
E-Mails: xliu.nk@gmail.com (X.L.); steanna@sohu.com (S.C.); liuhongler@ee.buaa.edu.cn (H.L.); husha89@gmail.com (S.H.)

[2] School of Computer Science, Nanjing Normal University, Nanjing 210097, China;
E-Mail: dqzhang@njnu.edu.cn

* Author to whom correspondence should be addressed; E-Mail: ninghuansheng@buaa.edu.cn.

**Abstract:** Sensing technology has been widely investigated and utilized for gas detection. Due to the different applicability and inherent limitations of different gas sensing technologies, researchers have been working on different scenarios with enhanced gas sensor calibration. This paper reviews the descriptions, evaluation, comparison and recent developments in existing gas sensing technologies. A classification of sensing technologies is given, based on the variation of electrical and other properties. Detailed introduction to sensing methods based on electrical variation is discussed through further classification according to sensing materials, including metal oxide semiconductors, polymers, carbon nanotubes, and moisture absorbing materials. Methods based on other kinds of variations such as optical, calorimetric, acoustic and gas-chromatographic, are presented in a general way. Several suggestions related to future development are also discussed. Furthermore, this paper focuses on sensitivity and selectivity for performance indicators to compare different sensing technologies, analyzes the factors that influence these two indicators, and lists several corresponding improved approaches.

## 1. Introduction

Recently, gas sensing, as a typical application in intelligent systems, is receiving increasing attention in both industry and academia. Gas sensing technology has become more significant because of its widespread and common applications in the following areas: (1) industrial production (e.g., methane detection in mines) [1–28]; (2) automotive industry (e.g., detection of polluting gases from vehicles) [29–43]; (3) medical applications (e.g., electronic noses simulating the human olfactory system) [44–60]; (4) indoor air quality supervision (e.g., detection of carbon monoxide) [61–73]; (5) environmental studies (e.g., greenhouse gas monitoring) [74–86].

During the last fifty years, different studies have established various branches of gas sensing technology. Among them, the three major areas that receive the most attention are investigation of different kinds of sensors, research about sensing principles, and fabrication techniques [87–115]. In this paper, a classification of sensing technologies is given, followed by descriptions of the main technologies to provide a comprehensive review. Two key performance indicators are highlighted, to introduce and compare different sensing technologies. Current research status and recent developments in the gas sensing field are reported, to discuss potential future interests and topics. Moreover, suggestions on related topics' future development are also proposed.

The remainder of the paper is organized as follows: Section 2 describes gas sensing methods' classification for further introduction, while Section 3 introduces performance evaluation and indicators of different gas sensing methods. Section 4 introduces sensing methods listed in Section 2 in detail. Technologies improving the sensitivity and selectivity of gas sensors are given in Section 5. Finally, Section 6 presents the conclusions.

## 2. Classification of Gas Sensing Methods

In order to give a clear introduction of sensing principles, this paper classifies gas sensing technologies into two groups: methods based on variation of electrical properties and other properties, as shown in Figure 1.

**Figure 1.** Classification of gas sensing methods.

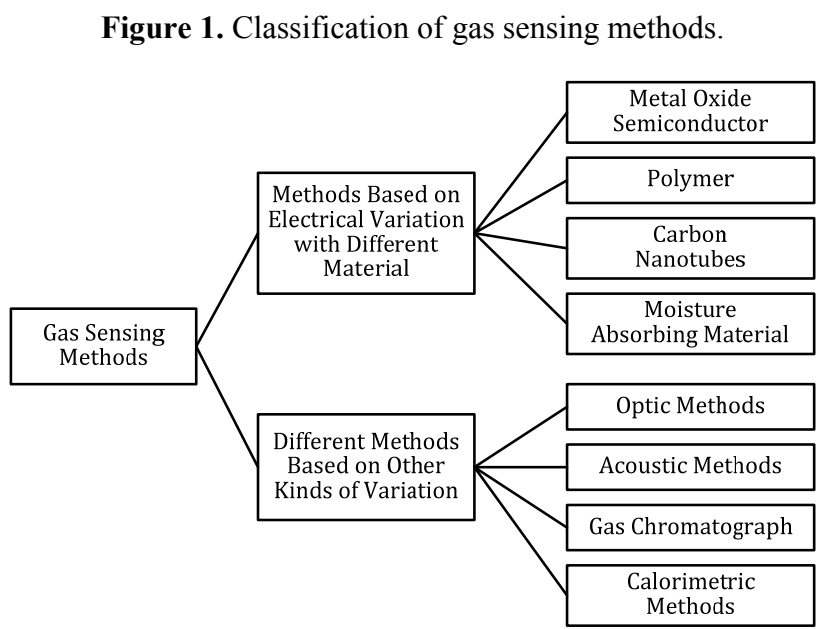

## 3. Performance Indicators and Gas Sensor's Stability

To evaluate the performance of gas sensing methods or gas sensors, several indicators should be considered: (1) sensitivity: the minimum value of target gases' volume concentration when they could be detected; (2) selectivity: the ability of gas sensors to identify a specific gas among a gas mixture; (3) response time: the period from the time when gas concentration reaches a specific value to that when sensor generates a warning signal; (4) energy consumption; (5) reversibility: whether the sensing materials could return to its original state after detection; (6) adsorptive capacity (also affects sensitivity and selectivity); (7) fabrication cost. More indicators for different sensing methods will be listed and compared in Section 4.

Besides, gas sensors designed for the market must guarantee the stability of their operation, in other words, they should exhibit a stable and reproducible signal for a period of time. There are several factors leading to gas sensor's instability (extracted and concluded from [116]): (1) design errors (which should be avoided); (2) structural changes, such as variations of grain size or grain network; (3) phase shifts, which usually refers to the segregation of additives doped with sensing materials; (4) poisoning triggered by chemical reactions; (5) variation of the surrounding environment. In order to solve these problems, the following methods could be considered: (1) using materials with chemical and thermal stability; (2) optimizing elemental composition and grain size of sensing materials; (3) utilizing specific technology during surface pretreatment of sensors. Among those indicators mentioned above, sensitivity and selectivity are the two main indicators, which will be discussed in Section 5.

## 4. Gas Sensing Methods

### 4.1. Methods Based on Variation of Electrical Properties

4.1.1. Metal Oxide Semiconductor

The most common sensing materials are metal oxide semiconductors, which provide sensors with several advantages such as low cost and high sensitivity. Generally, metal oxides can be classified into two types: non-transition and transition. The former (e.g., $Al_2O_3$) contains elements with only one oxidation state since much more energy is required to form other oxidation states, while the latter (e.g., $Fe_2O_3$) contains more oxidation states [81]. Therefore, transition-metal oxides could form various oxidation states on the surface, which is utilized by metal oxide semiconductors as sensing materials, compared to the non-transition ones. More precisely, transition-metal oxides with $d^0$ and $d^{10}$ electronic configurations could be used in gas sensing applications [99]. The $d^0$ configuration could be found in transition metal oxides (e.g., $TiO_2$, $V_2O_5$, $WO_3$), and $d^{10}$ appears in post-transition-metal oxides (e.g., $SnO_2$ and $ZnO$). Although most common metal oxide semiconductors sensitive to gas concentration are n-type semiconductors, there are also a few kinds of p-type semiconductors like $NiO_x$ (usually doped with n-type semiconductor like $TiO_2$) that could be used as gas sensor sensing materials. It is shown that 10% wt. $NiO_x$ content is needed to convert n-type conductivity into p-type. The main difference between n-type and p-type $NiO_x$ doped $TiO_2$ film is that as temperature increases, the sensitivity of n-type towards reducing gases is increased, while that of the p-type is decreased [117]. Therefore p-type semiconductors have relatively lower operating temperatures than n-type ones.

Sensors based on metal oxide semiconductors are mainly applied to detect target gases through redox reactions between the target gases and the oxide surface [118]. This process includes two steps [81]: (1) redox reactions, during which $O^-$ distributed on the surface of the materials would react with molecules of target gases, leading to an electronic variation of the oxide surface; and then (2) this variation is transduced into an electrical resistance variation of the sensors. The resistance variation could be detected by measuring the change of capacitance, work function, mass, optical characteristics or reaction energy [99].

Metal oxides, such as $SnO_2$, $CuO$, $Cr_2O_3$, $V_2O_5$, $WO_3$ and $TiO_2$, can be utilized to detect combustible, reducing, or oxidizing gases with sensors which are mainly based on the resistance change responses to the target gases [119]. Tin dioxide ($SnO_2$) *is* the commonly used gas sensing material. It is an n-type granular material whose electrical conductivity is dependent on the density of pre-adsorbed oxygen ions on its surface. The resistance of tin dioxide changes according to the variation of gas concentration (e.g., liquefied petroleum gas (LPG), methane ($CH_4$), carbon monoxide (CO) and other reducing gases [33,120], while the relationship between resistance and target gas concentration is nonlinear [68].) Other metal oxide semiconductors (e.g., tungsten trioxide ($WO_3$)) are also widely used for gas sensing. Anodic tungsten oxide applying electrochemical etching of tungsten shows excellent responses towards hydrogen ($H_2$) and nitrogen oxide (NO) [121]. However, the response of pure $WO_3$ to $NH_3$ is rather poor, and because of the interference from $NO_x$, the selectivity of $WO_3$ sensors for $NH_3$ is low. In order to implement $WO_3$ in gas sensing, $WO_3$ should be decorated with copper and vanadium as catalytic additives to improve the response, and the abnormal behavior of sensors should be eliminated [122]. Others like titanium dioxide ($TiO_2$) are also used as sensitive layers for their sensitivity in terms of dielectric permittivity to gas adsorption [123].

Several influencing factors, such as the characteristics and structure of the sensing layer, affect the redox reactions and thus decide the sensitivity of metal oxides as gas sensing materials. Among all sensors based on metal oxide semiconductors, the sensitivity of $SnO_2$-based ones is relatively high, leading to its greater popularity. However, this high sensitivity is mainly based on the high working temperature, which is often realized through a heated filament. For most metal oxide gas sensors, the high operating temperature is due to the reaction temperature of $O^-$ [99]. The sensitive layer has to be preheated to an elevated temperature in order to increase the probability of gas molecule adsorption on the layer surface which would consume ions of the sensing materials. As the ions are consumed, the conductivity of the film will increase to realize the sensing function. Besides the heated filament, the micro-hotplate is another choice for keeping the sensing materials at an elevated temperature [124]. Apart from the heating methods, there are also other methods like pre-concentration technology that could be applied to improve the sensitivity of gas sensors [69]. Since methods like those do not affect the characteristics of sensing materials, a detailed discussion will be given in Section 5. For the methods changing materials' characteristics, the use of composite materials such as $SnO_2$-$ZnO$ or $FeO_3$-$ZnO$ is also a good choice for improving metal oxide gas sensors' sensitivity since they suggest a synergistic effect between the two components [99]. In this method the sensitivity could be moderated by changing the proportions of each material in the composite.

The working temperature of $SnO_2$-based sensors is from 25 °C to 500 °C and the best sensing temperatures to various gases are different [14]. This could cause potential selectivity problems in applications, because if the temperature deviates too much from the optimal value, other gas components may be more reactive towards $SnO_2$, leading to poor selectivity. However, if the difference between these two temperatures is large, a single sensor could also be designed to detect two kinds of target gases at the same time. For example, the optimal sensing temperature of $SnO_2$ to $CH_4$ is 400 °C while that for CO is 90 °C, which requires a thermostatic cycle of the sensitive element at the two temperature values so that both gases could be detected by measuring the resistivity of the sensing element during each gas period [13] (as shown in Figure 2). The best sensing temperature of another common aggressive gas, hydrogen fluoride (HF), is different from the above (380 °C, but oxygen is more reactive toward $SnO_2$ than HF if the temperature is higher than 380 °C [14]. While the sensitivity of $SnO_2$ gas sensors could be controlled mainly by temperature, there are several methods that could improve the selectivity of the sensing materials. Selectivity could be improved by doping the sensing film surface with a suitable catalyst material [81,119]. Another common method to enhance the selectivity is using sensor arrays based on different sensing elements [79,125]. A gas sensor array is made up with two or more sensing elements in order to detect the gas with data of higher dimensions. For those with several elements, there is usually a gas recognition circuit for sensor arrays to enhance selectivity [125]. The more detailed introduction of sensor arrays is provided in Section 5. Besides, methods for improving the selectivity also include the use of catalytic filtering technology for combustible gases [126,127] and the compositional control of sensing materials (like $TiO_2$-$SnO_2$ composite, which also shows good sensitivity) [128].

**Figure 2.** A thermostatic cycle of a sensitive element for CO and $CH_4$.

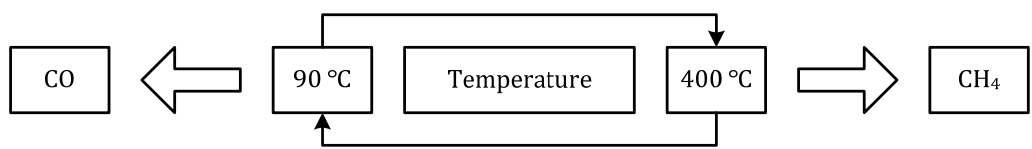

Sensors based on metal oxide semiconductors have been widely utilized. However, for some sensors, their demand of high operation temperature requires more cost and complicated configurations compared to others working at room temperature, which restricts their development. To solve this problem, researchers have come up with some methods such as the utilization of micro-sized sensor elements with micro-heaters fabricated by silicon IC technology [129] and temperature pulse operating mode with short heat intervals [130] which facilitates the operation of sensors with minimum power consumption. Another problem is the long recovery period needed after each gas exposure, which is not practical for some sensing devices like e-noses, and severely restricts their usage in applications where the gas concentrations may change rapidly. The structural instability and defects of other indicators also limit their field of application. In conclusion, faced with inherent challenges from their own nature and other kinds of gas sensors, research on sensors based on metal oxide semiconductors should find out some new solutions to overcome their defects. Studies about metal oxide nanostructures have shown that nanostructures (e.g., semiconductor nanowires) could improve gas sensors' sensitivity and response time [131].

4.1.2. Polymers

Generally, sensors based on metal oxide semiconductors exhibit significantly greater sensitivity to inorganic gases like ammonia and a few kinds of volatile organic compounds (VOCs) like alcohol ($C_2H_5OH$) and formaldehyde. However, some other VOCs which could cause adverse health effects when their concentration over a certain threshold cannot be detected by metal oxide semiconductor-based sensors. These VOCs could easily be breathed by humans since they are commonly used as ingredients in household products or in industrial processes where they normally get vaporized at room temperature. Therefore, it is important to monitor the concentration of these vapors to safeguard the health of residents and workers, and also to keep atmospheric emissions under control in order to avoid environmental hazards. Such exact requirements need sensing materials like polymers.

Although several studies consider polymer-based gas sensing materials applied in detecting inorganic gases like $CO_2$ and $H_2O$ [70], they are most frequently used to detect a wide range of VOCs or solvent vapors in the gas phase, such as alcohols, aromatic compounds or halogenated compounds. Like metal oxide semiconductors, when the polymer layers are exposed to the vapor of an analyte, the physical properties of the polymer layer such as its mass and dielectric properties will change upon gas absorption. Specifically, the various physisorption mechanisms by which VOCs interact with polymers include induced dipole/induced dipole interactions (also named London dispersion), dipole/induced dipole interactions, dipole/dipole interactions and hydrogen bonds (Lewis acidity/basicity-concept) [15]. In addition to the fact that mechanisms of these property changes are different from those of metal oxide semiconductors, the detection process can be expected to occur at room temperature. According to the different changes in physical properties, polymers used for gas sensing can be further classified into two groups: (1) conducting polymers, and (2) non-conducting polymers.

*Conducting Polymers*: it is well established that the electrical conductivity of these conducting polymers is affected through exposure to diverse organic and inorganic gases. This feature has led to the investigation of these materials as gas sensing layers by a number of groups [54,55,132]. Conducting polymers that can be used as gas sensing materials include polypyrrole (PPy), polyaniline (PAni), polythiophene (PTh) and their derivatives [96]. It should be highlighted that the conductivity of pure polymers is too low for them to function as gas sensing materials. Therefore, to realize the function of detecting gases, more work must be done. Previous studies showed that the conductivity of polymers can be improved through different processes of doping by redox reactions or protonation. After the doping process, which is reversible, polymers become conductors or semiconductors. The most important thing is that the doping level can be changed by chemical reactions between the polymers and target gases, making the detection of analytes based on these conducting polymers become practical. Most kinds of polymer are doped through redox reactions. For those gases inactive to redox reactions at room temperature, some specific polymers can be utilized. For example, redox reactions of CO could not happen at room temperature, however, the response of PAni towards CO could be observed [96]. Conducting polymers can be directly seen as transducers to reflect changes in electrical properties.

*Non-Conducting Polymers*: Non-conducting polymers have been widely utilized as sorptive coatings on different sensor devices, where the coating and the device can be regarded as a common transducer on the whole. Polymers with different properties or physisorption mechanisms can be coated onto respective transducers. For instance, polymer layers causing changes in resonance frequency, dielectric constant and enthalpy upon absorption/desorption of analytes can be respectively coated on mass-sensitive (e.g., Quartz Crystal Microbalance (QCM), Surface Acoustic Wave (SAW) and Surface Transverse Wave (STW)), capacitive (dielectric) and calorimetric sensor devices. Then, the sensor devices could convert changes in the monitored polymer properties into an electrical signal output [15]. Although the basic principles of non-conducting polymers in gas sensing are rather comprehensible, their performance is more complicated, even when coated on sensor devices of the same properties. For instance, STW resonant devices coated by thin sensitive polymer layers feature substantial advantages in terms of relative gas probing sensitivity, overall electrical performance and low noise of the sensor oscillator, compared to their SAW counterparts [133]. Additionally, non-conducting polymer membrane (e.g., polyimide) can also be used on metal oxide semiconductor gas sensors as molecular sieves, to enhance the overall selectivity by introducing the sensitivity of polymer layers [134].

Polymer-based gas sensors have advantages such as high sensitivities and short response times. Furthermore, while operation temperatures of metal oxide semiconductor-based sensors are usually more demanding, polymer-based sensors operate at room temperature. Therefore, the low energy consumption enables their applications in battery-driven detection units. Additional merits include the added benefit of low cost of fabrication, simple and portable structures, and the possibility of being reproducibly manufactured [53], by maintaining the conductive filler to insulate polymer ratio constant and dissolving the polymers into the solution in a uniform way.

Polymer-based gas sensors also have some disadvantages such as long-time instability, irreversibility and poor selectivity. The performance could also be affected by the working environment. Furthermore, the working principles of polymer as sensing materials still need a clearer and more convincing explanation. As a kind of gas sensors with lower power consumption polymer-based sensors have a promising future.

4.1.3. Carbon Nanotubes

Conventional sensing materials like metal oxide semiconductors have to deal with the problem of poor sensitivity at room temperature, while carbon nanotubes (CNTs) attract more attention because of their unique properties and have become the most promising materials for high-sensitive gas sensors. As a kind of promising sensing material, CNTs, have been found to possess electrical properties and are highly sensitive to extremely small quantities of gases, such as alcohol, ammonia ($NH_3$), carbon dioxide ($CO_2$) and nitrogen oxide ($NO_x$) at room temperature, while other materials like metal oxides have to be heated by an additional heater in order to operate normally. This high sensitivity eliminates the need of assisting technologies like pre-concentration, and thus contributes to the advantages of low cost, low weight and simple configuration. Besides, CNTs also outperform conventional sensing materials in term of great adsorptive capacity, large surface-area-to-volume ratio and quick response time, resulting in significant changes in electrical properties, such as capacitance and resistance [135]. Moreover, compared with metal oxide semiconductors that require microfabrication techniques, power supply and *ad-hoc* electronics when utilized as sensing material, CNTs possess good corrosion resistance and better bandwidth [136].

Generally, CNTs could be classified into single-walled carbon nanotubes (SWCNTs) and multiwall carbon nanotubes (MWCNTs). SWCNTs have been employed in RFID tag antennas for toxic gas sensing [82], in which the backscattered power from the tag antenna would be easily detected by the RFID reader if the concentration of ammonia rises to 4%, realizing the function of real-time gas sensing at room temperature. MWCNTs are usually used for remote detection of carbon dioxide ($CO_2$), oxygen ($O_2$), and ammonia ($NH_3$) depending on the measured changes in MWCN's permittivity and conductivity [83].

Like other gas sensing materials, the response time and property of CNTs are dissimilar for different target gases. The response property varies in physisorption and chemisorption. For example, it is shown that the response to $CO_2$ and $O_2$ is both linear and reversible, which means that the operation process only involves physisorption. However, the response to $NH_3$ is both irreversible and reversible, indicating both physisorption and chemisorption of $NH_3$ by the CNTs. Moreover, the response times to different gases are also different [83].

When utilized as sensing materials, on the one hand, CNTs are often decorated with other materials in order to enhance their sensitivity and selectivity. To enhance the selectivity to certain gases, CNTs could be mixed with silane which also raises CNTs' mechanical adhesion to the substrate [137]. In terms of improving sensitivity, oligonucleotides (DNA and RNA) could be utilized and the length of DNA sequence impacts the response process of CNT gas sensors [138]. On the other hand, CNTs could also be incorporated into other sensing materials such as metal oxide semiconductors to improve their sensitivity [122].

Another application of CNTs is the detection of partial discharge (PD) generated by decomposition of sulfur hexafluoride ($SF_6$). Partial discharge detection is an effective method to assess the insulation condition of gas-insulated switchgear (GIS). Partial discharge generated by the decomposition of $SF_6$ would lead to the increment of CNTs' electrical conductance, indicating the concentration of $SF_6$ in GIS [16]. Similarly, gas sensors based on graphene, another material composed by carbon elements have also shown good gas sensing characteristics, such as their whole volume is exposed to the surface adsorbate and few crystal defects, which maximizes the signal-to-noise ratio to a level sufficient for detecting variations of gas concentrations at room temperature [139]. It can be expected that novel materials like CNTs and graphene will attract more attention as gas sensing materials in the future.

4.1.4. Moisture Absorbing Material

Moisture absorbing materials could be embedded with RFID tags for detection of water vapor concentrations, namely the level of humidity, since their dielectric constant could be changed by the water content in the environment. If RFID tags are covered with moisture-absorbing materials like paper, absorbed water would lead to tag antenna near-field ohmic losses thus changing their resonance frequency, which could be detected by RFID readers. For passive RFID tags, the minimum power level offered by the RFID reader to power up the tags is determined by the water concentration, which could be expressed in term of the surrounding air's humidity. This procedure could be realized with only one tag, but the distance between tags and readers should be kept constant. An alternative method is utilizing two tags at the same time, of which one is covered with moisture-absorbing material and the other is untouched. In this method a table of the difference in power-up level *versus* the water concentration is needed [71]. Also mentioned in [71], sensors like that could be located inside walls or floors of buildings, mainly in order to prevent costly damage due to mold or decay and it could also be positioned under hidden water pipes for detection of leakage. Other applications detecting humidity levels, like water vapor concentration monitoring for food storage, could also utilize methods like those based on moisture absorbing materials and RFID tags.

Moisture absorbing materials could also be used as the RFID tag antenna's substrate rather than as a tag coverage since its dielectric constant could be controlled by the humidity of the surrounding air, which would cause the variation of antenna's performance that could be detected by the RFID reader. The moisture-absorbing material enveloped tags are low cost and suitable for mass production, thus more research in this area can be anticipated.

4.1.5. Classification Based on Sensors' Operating Modes

Apart from the proposed classification based on the mentioned sensing materials, sensors based on electrical variations could be divided into two types, according to the operating modes [135], one is based on direct contact, where the monitoring station and sensing unit are wired or could be wireless, enabling their integration with a wireless module (as shown in Figure 3(a)). The other type is based on wireless transducers utilizing the change of electrical indicators to indicate the variation of physical parameters, such as gas concentration. While most sensors belong to the first type, development of wireless transducers, especially high frequency ones, is very limited, requiring more extensive research. We think that the latter type of sensor could be regarded as a new type of reconfigurable antenna. While the reconfigurable section (like pin diodes or varactor) of reconfigurable antennas in the conventional sense is usually controlled by other devices like bias circuits, we could design the tags as a reconfigurable antenna without chips since the electric characteristics of sensing materials are reconfigurable (as shown in Figure 3(b)). Unlike conventional reconfigurable antennas, the reconfigurable part is controlled by the gas concentration.

**Figure 3.** Sensor system (**a**) integrated with wireless module and (**b**) based on wireless transducer.

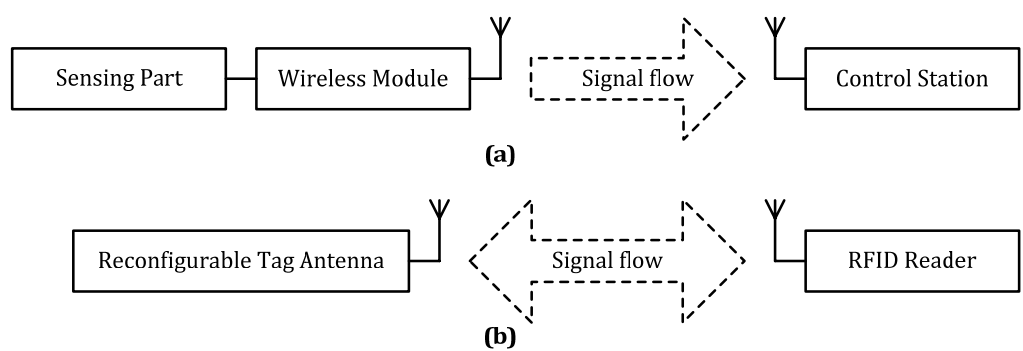

*4.2. Methods Based on Variation of Other Properties*

4.2.1. Optical Methods

Gas sensing by optical methods is usually straightforward and could achieve higher sensitivity, selectivity and stability than non-optical methods with much longer lifetime. Their response time is relatively short, which enables on-line real time detection. The performance will not be deteriorated by the changing environment or catalyst poisoning caused by specific gases, *etc*. Optical methods for gas sensing are mostly based on spectroscopy. However, their applications on gas sensors are seriously restricted due to miniaturization and relatively high cost. Only a few commercial gas sensors are based on optical principles. This paper will only give a general overview of optical gas analysis techniques, and illustrate a detailed introduction to off-the-shelf optical sensors.

Spectroscopic analysis mainly involves techniques based on absorption and emission spectrometry. The principle of absorption spectrometry is the concentration-dependent absorption (molar absorptivity ε) of the photons at specific gas wavelengths (*i.e.*, Beer-Lambert law). The precise wavelengths for specific gases can be found in the HITRAN database [140]. Apart from the basic method according to prior principles, there are many types of improved absorption spectrometry including Differential Optical Absorption Spectroscopy (DOAS) [34,72], Tunable Diode Laser Absorption Spectroscopy (TDLAS) [84,115,141], Raman Light Detection and Ranging (LIDAL) [35], Differential Absorption LIDAR (DIAL) [85,86], Intra-Cavity Absorption Spectrometry (ICAS) [142–144], *etc*. The law of emission spectrometry is that excited atoms will emit photons and then go back to the ground state. Laser-Induced Breakdown Spectroscopy (LIBS) [17] is one kind of technique based on emission spectrometry. Fourier Transform Infrared Spectroscopy (FTIR) [145] can be used in both absorption and emission spectrometry. Moreover, techniques like Photoacoustic Spectroscopy [146], and Correlation Spectroscopy [147] belong to spectroscopic analysis. Due to the mentioned factors, these techniques are more commonly applied to gas detectors, which allow for more complicated system design and higher cost to gain excellent sensitivity, selectivity and reliability [18], than gas sensors.

Infrared (IR)-source gas sensors based on optical sensing principles are widely used. According to the classification above, IR-source gas sensors are based on the basic absorption spectrometry, and more specifically, the principle of molecular absorption spectrometry.

It means that every gas has its own absorbing property to IR radiation with different wavelengths, and thus its unique IR absorption fingerprint. As illustrated in Figure 4(a), an IR-source gas sensor contains three major parts: IR source, gas chamber and IR detector. When the IR source emits broadband radiation including the wavelength absorbed by the target gas, the sample gas in the gas cell will absorb the radiation in its particular way. The optical filter is used to screen out all radiation except for the wavelength that is absorbed by the target gas. Therefore, the presence of interested gas could be detected and measured by an IR detector. This system is also known as Non-Dispersive Infrared (NDIR) gas sensor.

Apart from the single detector mode, another cell containing reference gas could be added to enhance the accuracy of the IR sensing technique by eliminating the ambient environmental factors while not increasing the complexity too much. The two-detector layout is shown in Figure 4(b). In order to guarantee that the radiation parameters of IR source are identical, the two gas cells can use the IR waves reflected by the same source by using the property of mirror as in the picture.

Among all three components mentioned above, the selection of the wavelength range of IR source has a great influence on the final detection result. The mid-infrared spectral region is usually of interest because it supports stronger molecular absorption than near-infrared does. However, traditional mid-IR laser sources have some major drawbacks: lack of continuous wavelength tenability, low output power and cooling requirement of lead salt diode lasers, complexity and low power of nonlinear optical sources. In order to combat the above disadvantages, many new lasers have been investigated, like Quantum-Cascade Lasers (QCLs) providing tunability in the spectroscopically important region from 3 to 20 μm, excellent properties in terms of narrow line width, low average power (tens of milliwatts) and room temperature operation [48].

**Figure 4.** IR-source gas sensors (**a**) based on the basic absorption spectrometry and (**b**) with reference filter/detector.

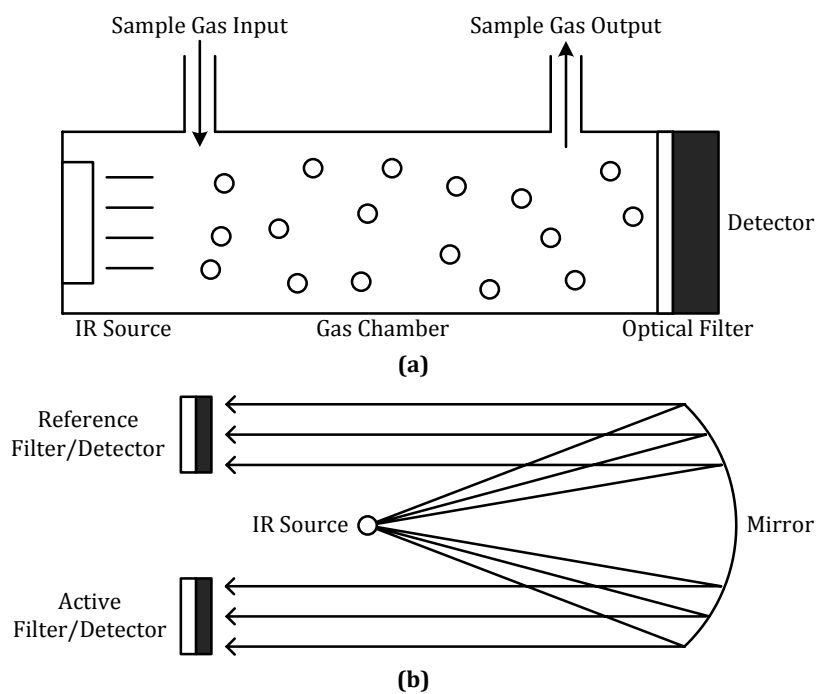

Among all frequency bands, there are two reasons for the particular attention paid by researchers and manufacturers to the IR frequency band. First, the measurement range of other frequency bands will be reduced in the presence of steam, mist or smoke due to scattering, additional absorption and diffraction. But IR radiation is less scattered by particles of smoke or mist than visible radiation because of its longer wavelength [19]. Second, MEMS technology enables the miniaturization of infrared sources and detectors, which greatly contributes to the popularization of IR sensors.

The application potential of IR-based and NDIR-based sensors has already been verified in areas like remote air quality monitoring and gas leak detection systems with high accuracy and safety [19,73]. It is predictable that these sensors will occupy the high-end market with their increasingly extraordinary performance.

4.2.2. Calorimetric Methods

Pellistors constitute a major class of electrical gas sensors which are calorimetric in nature. They are solid-state devices used to detect either combustible gases or those having a significant difference in thermal conductivity compared to that of air [148]. The detecting elements consist of small "pellets" of catalyst-loaded ceramic whose resistance varies in the presence of target gases, and hence the term "pellistor" being a combination of "pellet" and "resistor". The limit of detection (LOD) for calorimetric sensors is typically in the low parts-per-thousand (ppth) range (e.g., VQseries pellistors from e2v

Technologies) [148], which is suitable for industrial scenarios, but obviously not enough for laboratory applications.

Specifically, pellistors can be divided in two types: Catalytic and Thermal Conductivity (TC). In each case, some characteristic property of the gas (combustion enthalpy or thermal conductivity, respectively) creates a temperature variation that will be subsequently measured resistively, either by a platinum resistance temperature detector or a thermistor.

The catalytic sensors measure the heat evolution from catalytic oxidation of the gas analyte, and are the most common commercial type of pellistors. They work by burning the target gas and generating a specific combustion enthalpy, thus enabling the detection of low concentration analytes in a short response time [20,37].

As the successor of the flame safety lamp for detecting combustible gases, the catalytic sensors increase the level of accuracy and enable the incorporation into detection systems. But they still suffer from a serious problem-catalyst poisoning in the presence of some specific impurities in gas samples, which will lead to drastically reduced and sometimes irreversible loss of catalyst activities.

Figure 5(a) is the schematic diagram of a catalytic sensor. A high surface area catalytic layer is laid down on a ceramic bead, as shown in Figure 5(b), which contains a platinum coil acting as the heater/calorimeter. The platinum coil is heated until the catalytic layer is at <500 °C, and the combustible gas will burn on the surface of this catalytic layer. The generated heat produces a change in the resistance of the coil, and then could be measured by simple circuits, such as a Wheatstone bridge circuit [149].

**Figure 5.** Catalytic sensor (**a**) schematic diagram and (**b**) configuration of ceramic bead.

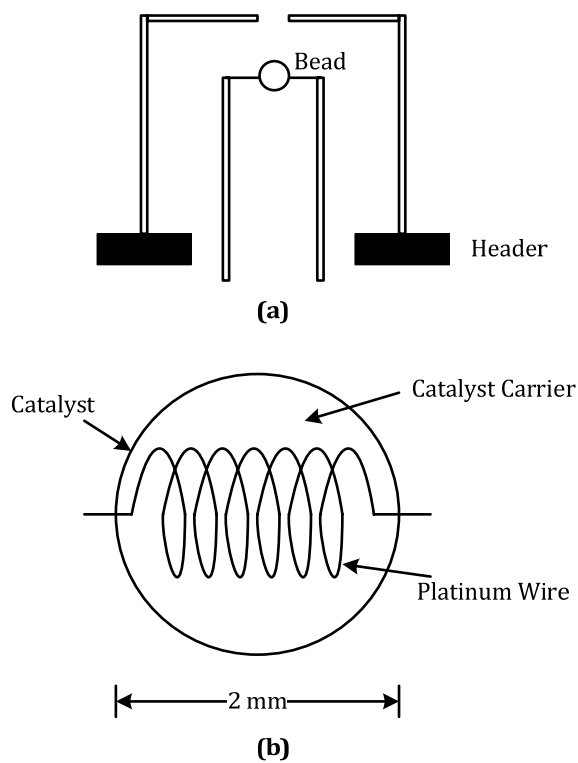

Studies also showed that catalytic sensors could get proper output sensitivity at lower temperatures with the assistance of electric fields [150]. The low temperature could be a merit, because the surface decrease of catalytic sensors under this scenario is slow and the output stability will be improved. Moreover, it was shown that if catalytic sensors were pre-poisoned in an electric field, the process of attenuation of sensitivity in poisoning experiments was slower than that of ordinary catalytic sensors.

The TC sensors rely on the measurement of heat dissipation into gas analytes, or in other words, gases are identified based on their thermal conductivity [36]. The target gas is pumped into a gas chamber, where the heater/calorimeter (e.g., platinum or tungsten wire, thermistor) is in the center. This center component is heated to a specific temperature, and then the resistance of the component provides information about the thermal conductivity of the gas and hence the identity of the gas. As the application, piezoresistive cantilevers in gaseous environments with potentially varying thermal conductivity, temperature, and flow rate, have a dominating response based on thermal conductivity changes [151]. TC sensors have a large detection range with the highest detection concentration near 100%. They also have advantages like good stability, reliability, and simple equipment. But their accuracy and sensitivity need to be further enhanced.

There are also some other calorimetric methods besides the two mentioned above, such as methods based on enthalpy change, providing that chemical reactions or physisorption processes all release or absorb heat from surroundings. Apart from by resistance temperature detectors and thermistors, temperature differences could also be measured by thermocouples, which make use of the Seebeck-effect [125].

Calorimetric sensors generally suffer from a deficiency in selectivity, which originates from their inherent physical mechanisms: many pure gases (or mixtures thereof) may have similar combustion enthalpies or thermal conductivities, but it is noteworthy that numerous industrial applications of commercial calorimetric sensors are based on the prior knowledge that the gas constituents are known or in limited numbers, and have sufficiently dissimilar physical characteristics. Therefore, the sensors can be appropriately selected and calibrated for the analytes to be monitored.

To improve the performance of existing calorimetric sensors for their industrial applications, future attention need to be paid in the following areas: (1) Reducing the power consumption; (2) Increasing the sensors' resistance to poisoning and mechanical shock; (3) Designing "flameproof" sensors, which involve an enclosure enclosing the beads to prevent the ignition of the gas around the sensing beads to be transmitted to the bulk of gas so as to prevent explosions.

4.2.3. Gas Chromatograph

Gas Chromatography (GC) is also a common method for gas sensing, but more accurately, it is a typical laboratory analytical technique which has excellent separation performance, high sensitivity and selectivity [152]. GC includes substantial quantitative analytical methods, for example, Volatile Sulphur Compounds (VSCs) could be analyzed by Flame Photometric Detection (FPD), Pulse Flame Photometric Detection (PFPD), Sulphur Chemiluminescence Detection (SCD), and Atomic Emission Detection (AED) [21]. GC-Olfactometry (GC-O) method [153] which combines human perception of odor and chromatographic separation of compounds is also a branch of GC. However, the cost of GC is high, and its miniaturization for portable application needs more technological breakthroughs. Therefore, GC does not quite satisfy the device and material constraints for unattended, flexible basic sensors.

4.2.4. Acoustic Methods

Gas sensors based on chemical principles experience some intrinsic weaknesses that are difficult to overcome, especially when applied in Wireless Sensor Networks (WSNs), like short lifetime and secondary pollution which could be exactly avoided by ultrasonic methods [154]. Measurement parameters based on ultrasonic methods mainly fall into three categories: (1) speed of sound; (2) attenuation; and (3) acoustic impedance.

Measuring sound velocity is the best studied category. The major sound velocity detection method is Time-of-Flight (TOF), which uses the travel time of ultrasoound at a given distance (*i.e.*, sound distance) to calculate the propagation velocity of ultrasonic waves.

Figure 6 [24] illustrates two channels, which are exactly alike, respectively measuring ultrasonic propagation parameters (e.g., time difference $\Delta t$ or sound wave phase) in the reference gas and the gas mixture of interest. The gas concentration is determined by difference methods [22,23].

**Figure 6.** Method of ultrasonic detection.

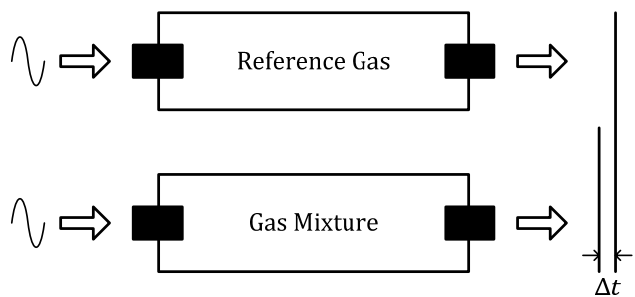

This conventional ultrasonic detection method has high precision, but it is hard to guarantee that the impacts of environment on the two channels are identical. Because of the large power consumption, it is also difficult to build into wireless nodes for ultrasonic gas concentration detection. Therefore, researchers have previously developed an improved TOF for binary mixtures. It uses a single-channel for gas mixtures, and uses the relationship between the speed of sound in the air and the air temperature to determine the standard of reference speed of sound $C_R$, and hence the time difference $\Delta t$ between transmitting in the air and target gases [24].

The measured gas velocity can be used to determine many gas properties, such as the concentration of the target gas, based on the mathematical reasoning that it is proportional to the time difference of sound propagation [24], to identify a special kind of gas through which the sound velocity is different from others in the group [38], and to calculate the composition or the molar weight of different gases in mixtures based on some thermodynamics equations [155–157].

Attenuation refers to the energy lost as thermal or scattered energy when an acoustic wave travels through a medium [158]. Different gases exhibit different attenuation properties, hence providing a way to detect specific gases. Attenuation could also be used in combination with sound speed to determine gas properties [159]. Compared with the sound speed method, attenuation is a relatively less robust method, which is inclined to be influenced by turbulence, particles or droplets in the gas, or even transducer degradation overtime. This may be the reason why hardly any commercial gas sensors are based on acoustic attenuation.

Acoustic impedance is usually used to determine gas density, since the acoustic impedance is given by the simple equation: $Z = \rho C$, where $\rho$ is the gas density and $C$ is the sound speed. Thus, by the measured acoustic impedance and sound speed, the density of a gas could be calculated [160]. However, the measurement of gas acoustic impedance is extremely difficult in practice, especially in a process environment. Therefore, like attenuation, acoustic impedance has not been widely applied on commercial acoustic sensors.

*4.3. Summary and Comparison for Different Gas Sensing Methods*

Table 1 gives an overall summary in advantages, disadvantages and application fields for the mentioned gas sensing methods.

**Table 1.** Summary of basic gas sensing method.

| Materials | Advantages | Disadvantages | Target Gases and Application Fields |
|---|---|---|---|
| Metal Oxide Semiconductor | (a) Low cost; (b) Short response time; (c) Wide range of target gases; (d) Long lifetime. | (a) Relatively low sensitivity and selectivity; (b) Sensitive to environmental factors; (c) High energy consumption. | Industrial applications and civil use. |
| Polymer | (a) High sensitivity; (b) Short response time; (c) Low cost of fabrication; (d) Simple and portable structure; (e) Low energy consumption. | (a) Long-time instability; (b) Irreversibility; (c) Poor selectivity; | (a) Indoor air monitoring; (b) Storage place of synthetic products as paints, wax or fuels; (c) Workplaces like chemical industries. |
| Carbon Nanotubes | (a) Ultra-sensitive; (b) Great adsorptive capacity; (c) Large surface-area-to-volume ratio; (d) Quick response time; (e) Low weight. | (a) Difficulties in fabrication and repeatability; (b) High cost. | Detection of partial discharge (PD) |
| Moisture Absorbing Material | (a) Low cost; (b) Low weight; (c) High selectivity to water vapor. | (a) Vulnerable to friction; (b) Potential irreversibility in high humidity. | Humidity monitoring |
| Optical Methods | (a) High sensitivity, selectivity and stability; (b) Long lifetime; (c) Insensitive to environment change. | (a) Difficulty in miniaturization; (b) High cost. | (a) Remote air quality monitoring; (b) Gas leak detection systems with high accuracy and safety; (c) High-end market applications. |
| Calorimetric Methods | (a) Stable at ambient temperature; (b) Low cost; (c) Adequate sensitivity for industrial detection (ppth range). | (a) Risk of catalyst poisoning and explosion; (b) Intrinsic deficiencies in selectivity. | (a) Most combustible gases under industrial environment (b) Petrochemical plants; (c) Mine tunnels; (d) Kitchens. |
| Gas Chromatograph | (a) Excellent separation performance; (b) High sensitivity and selectivity. | (a) High cost; (b) Difficulty in miniaturization for portable applications. | Typical laboratory analysis. |
| Acoustic Methods | (a) Long lifetime; (b) Avoiding secondary pollution. | (a) Low sensitivity; (b) Sensitive to environmental change. | Components of Wireless Sensor Networks. |

## 5. Some Approaches to Improve Sensitivity and Selectivity

In many cases, sensitivity and selectivity are both vital indicators in designing and applying sensors. For instance, since the concentration of some toxic gases in the ambient environment should not surpass a particular value, sensors should have sufficient sensitivity to give people a signal before the concentration reaches a dangerous value. Moreover, since sensors' target gases are usually mixed with others, the selectivity level of gas sensors which determines whether they could detect the target gases without being interfered by other ones is also quite important. Therefore, sensitivity, selectivity and the corresponding approaches are presented as follows.

*5.1. Introduction for Sensitivity and Improving Approaches*

Usually, sensitivity could be defined as the ratio of $R_a$ to $R_g$ for reducing gases or the reciprocal for oxidizing gases, where $R_a$ stands for the resistance of gas sensors in the reference gas and $R_g$ stands for the one in the reference gas containing the target gases [99]. It could also be measured by the minimum value of target gases' volume concentration when the gas sensors could discover the target gases, and thus obviously, the lower the value, the better the sensitivity of the gas sensors. Different levels of sensitivity are required under different circumstances, for example, ppth (parts per thousand) is enough for rough industrial detection, while low ppm (parts per million) or ppb (parts per billion) are needed for more precise monitoring in laboratories. As mentioned in Sections 2 and 3, there are several factors determining the sensitivity of gas sensing methods. Chemical components of sensing materials, physical structure of sensing layers, humidity and temperature of the environment are the potential factors that people could study to improve the sensitivity. In the following, some examples are given.

Firstly, in term of sensing materials' structure, the dielectric resonator is a good choice for gas sensing because of its large surface area and sensitivity to the external environment. The relative permittivity of the dielectric resonator changes according to the concentration of the target gas, based on which dielectric resonators function as gas sensor. However, it should be of concern that common cylindrical dielectric resonators in millimeter-wave frequency band become difficult to machine because their dimensions are impractically small when used in their conventional TE or TM modes [161]. Therefore, the dielectric resonator operating with Whispering Gallery Modes (WGM) could be used in gas sensor [123]. The one with WGM is described as comprising waves running against the concave side of the cylindrical boundary of the rod and the waves move essentially in the plane of the circular cross section, so most energy is confined between the cylindrical boundary and an inner modal caustic [161]. Compared to the conventional way of excitation, the one with WGM has more advantages such as large dimensions millimeter-wavelength band and high level of integratability. In addition, since they could operate in the azimuth direction, gas sensing in all directions could be realized.

Secondly, as shown in Section 4, for some metal oxide semiconductor-based sensing methods, the temperature of environment is a key parameter guaranteeing the normal operation of gas sensors, which is usually realized by heaters. While one sensor is designed to detect two or more kinds of gases simultaneously, the sensitivity of all gases to be detected should be improved. Then a thermostatic cycle is needed to ensure the sensitivity of the sensor since different gases have different best sensing temperatures. In the thermostatic cycle of two temperature values, sensors measure the resistivity of the sensing element during each gas period to ensure that the sensor is able to monitor the concentration of both two gases accurately and simultaneously [13]. Simply, it is applicable to control the temperature of sensing materials to a desired value through a filament. Different from those sensors designed for one certain kind of gas, sensors with a thermostatic cycle should have a cooler in order to turn the value of temperature to the lower one instantly. It should be mentioned that both of the gas periods should be quite short since the process of gas sensing is in real time. Therefore, the process of detecting two-kind gases could be combined into a single element. If sensors with thermostatic cycle of several elements are designed, like sensors detecting three kinds of gases or more, the temperature-control structure will become more complex.

Thirdly, for gases which are toxic or combustible even at a small concentration levels, sensitivity of sensing methods should be designed as a very high value. However, we notice that gas sensing in low ppb level is a relative challenging task for gas sensors, since the LOD may not be able to be improved due to the intrinsic sensing mechanism or sensitivity of materials. For instance, in [69], the authors announced that the evaluation result from their laboratories showed that commercial off-the-shelf sensors could not give a reliable reading when the formaldehyde concentration is 10 ppb. Pre-concentration, however, is a very promising technique combined with cryogenic trapping, adsorption on porous polymers, molecular sieves and so on. The pre-concentrator could also act as a front-end detector device, and significantly enhance the detection level. The pre-concentration system basically operates in three steps: absorption, desorption and conditioning. In the absorption period, the testing gas is pumped into the pre-concentrator for the required period of time, and the target gas going through the absorbent is trapped/absorbed, while normal air passes through the absorbent (as shown in Figure 7). In desorption period, the pre-concentration tube is heated to a specific temperature to release the gas collected by the absorbent. Then the desorption flow is open and enables the sensor to detect the concentrated gas. For the last conditioning period, the tube is cooled to room temperature for the next gas testing cycle. The performance of the pre-concentrator is partially measured by a pre-concentrator factor, which means the ratio between concentration of input and output gas of the pre-concentrator. It is positively correlated to surface-area-to-volume-ratio (SA/V). A higher SA/V leads to a higher pre-concentrator factor, and hence the better pre-concentration performance to some extent [162]. In fact, pre-concentration performance is also related to other factors, like pressure drop, desorption rate, *etc.* [163–165]. By far, there are many kinds of pre-concentrators available. For example, solid trap/thermal desorption based pre-concentration system for the formaldehyde gas [69], resulting in the concentrator factor greater than 40 times when the formaldehyde concentration is less than 100 ppb, which makes the detection of trace amount of formaldehyde in indoor environments feasible. Several MEMS gas pre-concentrators (μGPC) are proposed for toxic gas sensing, which demonstrate an overall improved selectivity and detection limit (10 ppb or below) [166].

**Figure 7.** Absorption period of pre-concentration technology.

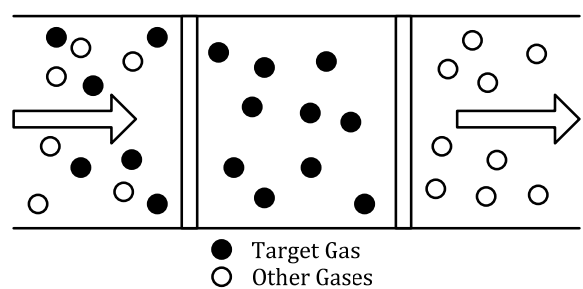

● Target Gas
○ Other Gases

Finally, it cannot be ignored that sensitivity of sensing methods based on other properties is relatively higher than that of the ones based on electrical variation. Therefore, research on approaches improving these kinds of methods' sensitivity also have a profound meaning. A method named Photoacoustic Spectroscopy which combines optics and acoustics improving sensitivity is introduced as follows. Photoacoustic Spectroscopy (PAS) is a widely used method for trace gas sensing through the adsorption of infrared or light radiation by the gas molecules inside the photoacoustic cell. The process of absorption of modulated light of appropriate wavelength by molecules leads to an increased gas temperature and a rise of the pressure in the cell, resulting in the generation of an acoustic wave whose amplitude is directly proportional to the gas concentration and can be detected using a sensitive microphone if the laser beam is modulated in the audio frequency range [48]. The microphones could be classified into three kinds: electret, capacitive and optical cantilever. Compared with the former two kinds, an optical cantilever microphone offers higher responsivity [146]. In a cantilever microphone system, pressure difference between the non-absorbing gas and target gas will decide the position of the cantaliver, which could be detected by sensors. The concentration of the target gas will be obtained through the cantilever's position and related equations.

*5.2. Introduction for Selectivity and Improving Approaches*

Selectivity is the ability of gas sensors to verify a specific kind of gas among gas a mixture; to distinguish ammonia ($NH_3$) from nitrogen oxides ($NO_x$) is a common example. Gas sensors usually have the drawback of poor selectivity, causing them to respond, not necessarily uniquely, to multiple analytes, which is also known as cross-sensitivity. The responses from a cross-sensitive sensor are therefore intrinsically ambiguous, and cannot serve as unique identifiers or signatures for gas samples of completely unknown composition [167]. Generally, approaches that improve selectivity fall in two categories. One is exploiting different properties of target gases to give them multi-dimensional signatures. Even if the selectivity on some properties may be poor when mixed with specific gases, the cross-sensitivity with one specific interfering gas does not exist on all properties, and thus the overall sensing result could distinguish target gases with a far more precise performance. The other is utilizing the difference between optimal conditions for sensing target gases and corresponding conditions for other gases. A straightforward method is to provide the gas mixture with the optimal condition for the target gas, by actual environment control or by compensation and offset, to achieve the best sensitivity for the target gas, while worsening the sensitivity of other gases. For the first kind of methods, utilization of sensor arrays is the main method to solve the selectivity problem. The concept of sensor arrays (sometimes known as multi-modality [168]) was examined in [169]. This approach employs different cross-sensitive sensing elements, possibly with different sensing principles, to achieve selective detection. Given the sensing responses are sufficiently different from each other (orthogonal is more important than substantial [15]), in the multidimensional space, it could provide a more distinct signature. Combined with pattern recognition, the analysis of gas mixtures could be realized.

For example, three transducers: mass sensitive, calorimetric and capacitive could form an array, which responds to fundamentally different analyte properties to detect volatile organic compounds. For the mass sensitive transducer, the absorption of analyte causes shifts in resonance frequency as a consequence of changes in the oscillating mass. The second calorimetric transducer detects enthalpy changes upon absorption (*i.e.*, heat of condensation) or desorption (*i.e.*, heat of vaporization) of analyte molecules. The capacitive transducer monitors change in the dielectric constant of the polymer upon absorption of the analyte into the polymer matrix [15]. There are several fundamental requirements for sensors in a sensor array [169]:

(1) Sensors are functionally stable and reliable;
(2) Sensors respond to different kinds of gases, *i.e.*, sensors possess the cross-sensitivity to some extent. This could help reduce the amount of sensors and increase the array efficiency;
(3) The response and recovery time should be short, which reflects the detection efficiency of sensors. Otherwise, sensors will suffer from errors for being unable to follow the changes of gas composition and concentration in the gas chamber.

Apart from increasing the dimensions of gas signatures, the exploitation of sensor arrays has some extended functions to increase the selectivity or accuracy of gas sensors, such as damping cross-sensitivity at special situations and eliminating environmental impacts. For instance, multiple sensors could be tuned to detect a specific gas in a mixture by heating it to the temperature of maximum sensitivity for that gas, and thus avoid the cross-sensitivity at some temperature point [170]. On the other hand, sensor arrays could also help reduce the effects from the environment, such as temperature and humidity, by calibration using various responses [83]. Moreover, the concept of sensor array can also be extended into the RFID field [71]. Ordinary RFID tags are exploited as remotely read moisture sensors incorporated into one label. The common method is to embed one of the tags in a moisture absorbent material and leave the other one open. Thus, the level of relative humidity is determined by a passive RFID system by comparing the difference in RFID reader output power required to power up respectively the open and embedded tag.

Besides sensor array, other methods could also be utilized to improve selectivity, especially for the gases reacting quite differently under different conditions. For example, the thermostatic cycle designed for two specific gases just utilizes sensing materials' selectivity in different temperatures. At one temperature, sensitivity of one gas is improved to the maximum value while others' are reduced. Pre-concentrator could also be utilized to improve selectivity if only the target gas is concentrated selectively.

*5.3. Summary of Approaches Improving Sensitivity and Selectivity*

Table 2 gives an overall summary for corresponding approaches improving sensitivity and selectivity mentioned above. It should be noticed that approaches mentioned above are the ones that do not change the sensing materials' own configuration. There are also some methods that modify sensing materials' configuration or composition like decorating them with catalytic metal nanoparticles [171–173] or compositional control mentioned in Section 4, which could also improve gas sensors' sensitivity and selectivity.

Table 2. Summary of approaches improving sensitivity and selectivity.

| Approaches | For Sensitivity | For Selectivity |
|---|---|---|
| Dielectric Resonator | (a) large surface area<br>(b) permittivity changes with target gas' concentration<br>(c) WGM used in gas sensor | N/A |
| Thermostatic Cycle | (a) guarantee the best sensitivity of all target gases in each gas period<br>(b) for gases with quite different sensing temperature | (a) guarantee the best selectivity of all target gases in each gas period<br>(b) for gases with quite different sensing temperature |
| Pre-concentrator | (a) relative concentration of target gases is improved | (a) if pre-concentrator is selective |
| Photoacoustic Spectroscopy | (a) combine advantages of both optic and acoustic methods | N/A |
| Sensor Array | N/A | (a) provide with multi-dimensional signatures<br>(b) for gases with different sensing conditions, the difference is either large or small |

## 6. Conclusions

A survey on gas sensing technologies has been presented. This paper first gives the classification of gas sensing technologies based on variation of electrical or other properties, and then comprehensively reviews sensing principles and the typical characteristics of the various types of sensors. Considering two key performance indicators-sensitivity and selectivity-different kinds of sensing technologies are compared and evaluated. Moreover, factors that affect the sensing performance are analyzed, and pertinent methods for improvement are discussed. The outlook of future development in gas sensing field is also mentioned through this review. It is concluded that studies on gas sensing technologies should concentrate more on solving urgent problems like high energy consumption and fabrication complexity.